\documentstyle[aps]{revtex}
\begin{document}
\draft
\title{Maximum bounds on the surface redshift of anisotropic stars}
\author{B.V.Ivanov\thanks{%
E-mail: boyko@inrne.bas.bg}}
\address{Institute for Nuclear Research and Nuclear Energy,\\
Tzarigradsko Shausse 72, Sofia 1784, Bulgaria}
\maketitle

\begin{abstract}
It is shown that for realistic anisotropic star models the surface redshift
can not exceed the values $3.842$ or $5.211$ when the tangential pressure
satisfies the strong or the dominant energy condition respectively. Both
values are higher than $2$, the bound in the perfect fluid case.
\end{abstract}


\section{Introduction}

It is well known that the surface redshift $s$ for a static perfect fluid
sphere, whose density is positive and not increasing outwards, is not larger
than $s_m=2$ \cite{one}. The bound holds for the interior Schwarzschild
solution with constant fluid density $\rho $ and infinite central pressure.
It occurs before the appearance of a horizon.

During the years, different arguments have been put forth for the existence
of anisotropy in star models. It may be due to the presence of solid core,
phase transitions, mixture of two fluids, slow rotation \cite{two}. The idea
that the tangential pressure $t$ may be different from the radial pressure $%
p $ was suggested first by Lemaitre in 1933 \cite{three}. He discussed a
model sustained solely by tangential pressures and with constant density.
This model was generalized for variable density by Florides \cite{four} and
was recently revisited in Ref. \cite{five}. The application of anisotropic
fluid models to neutron stars began with the pioneering work of Bowers and
Liang \cite{six} and was done both analytically and numerically \cite
{seven,eight,nine,ten,eleven,twelve,thirteen,fourteen,fifteen}. Some recent
work may be found in Refs. \cite{sixteen,seventeen,eighteen,nineteen,twenty}.

In many papers it is stressed that arbitrarily big redshifts are obtained
when $t$ grows to infinity \cite{five,six,seven,ten,twelve,sixteen}.
However, in realistic models $t$ should be finite, positive, and should
satisfy the dominant energy condition (DEC) $t\leq \rho $ or even the strong
energy condition (SEC) $2t+p\leq \rho $. These may be written together as $%
t\leq \varepsilon \rho $ where $\varepsilon =1$ for DEC and $\varepsilon
=1/2 $ for SEC (if the realistic condition for positive $p$ in the interior
is accepted).

The bounds on $s$ in anisotropic models were studied in Ref. \cite{twentyone}
in a more general setting which incorporates soap bubbles, monopoles and
wormholes and the focus was that a horizon does not form. It generalized
this result in the perfect fluid case \cite{twentytwo} to anisotropic
fluids. Similar conclusions were reached in Ref. \cite{twentythree}.

In this paper we elaborate further on the method used in Ref. \cite
{twentyone} and give concrete values for the maximum surface redshift when
the tangential pressure satisfies either DEC or SEC.

In Sec. II the field equations are given in a convenient form and the bound
on $t$ is implemented to derive an inequality for the mass-radius ratio.

In Sec. III this inequality is used to obtain maximum bounds on the surface
redshift when SEC, DEC or a pseudoisotropy condition is satisfied by $t$.
Several realistic models are discussed and it is shown that the bounds can
not be saturated.

Sec. IV contains short discussion.

\section{Field equations and the main inequality}

The metric element in curvature coordinates is

\begin{equation}
ds^2=e^\nu d\tau ^2-e^\lambda dr^2-r^2\left( d\theta ^2+\sin ^2\theta
d\varphi ^2\right) ,  \label{one}
\end{equation}
where $\nu ,\lambda $ depend only on the radial coordinate $r$. The Einstein
equations read \cite{eight} 
\begin{equation}
\kappa \rho =\frac{e^{-\lambda }}r\left( \lambda ^{\prime }-\frac 1r\right) +%
\frac 1{r^2},  \label{two}
\end{equation}
\begin{equation}
\kappa p=\frac{e^{-\lambda }}r\left( \nu ^{\prime }+\frac 1r\right) -\frac 1{%
r^2},  \label{three}
\end{equation}
\begin{equation}
\kappa t=\frac 12e^{-\lambda }\left( \nu ^{\prime \prime }-\frac{\lambda
^{\prime }\nu ^{\prime }}2+\frac{\nu ^{\prime 2}}2+\frac{\nu ^{\prime
}-\lambda ^{\prime }}r\right) ,  \label{four}
\end{equation}
where $^{\prime }$ means derivative with respect to $r$, $\kappa =8\pi G/c^4$
and we use units with $G=c=1$. Eq. (2) integrates to 
\begin{equation}
z\equiv e^{-\lambda }=1-\frac{2m}r,\quad m=\frac \kappa 2\int_0^r\rho r^2dr.
\label{five}
\end{equation}
Here $m$ is the mass function, $m\geq 0$ and consequently $z\leq 1$. Eqs.
(3)-(4) give a linear second-order equation for $y\equiv e^{\nu /2}$: 
\begin{equation}
2r^2zy^{\prime \prime }+\left( r^2z^{\prime }-2rz\right) y^{\prime }+\left[
2\left( 1-z\right) +rz^{\prime }-2\kappa \Delta r^2\right] y=0.  \label{six}
\end{equation}
Here $\Delta =t-p$ is the anisotropy factor. Finally, Eq.(3) may be written
as 
\begin{equation}
y^{\prime }=\frac{1-z+\kappa pr^2}{2rz}y.  \label{seven}
\end{equation}
For stability reasons $p$ must be positive and hence $y^{\prime }>0$. Eq.
(6) has a more compact form 
\begin{equation}
z^{1/2}\left( \frac{z^{1/2}y^{\prime }}r\right) =Dy,\quad D=\left( \frac m{%
r^3}\right) ^{\prime }+\frac{\kappa \Delta }r.  \label{eight}
\end{equation}
It is convenient to introduce the average density 
\begin{equation}
<\rho >=\frac 3{r^3}\int_0^r\rho r^2dr=\frac{6m}{\kappa r^3}.  \label{nine}
\end{equation}
Then Eqs. (5), (8), (9) give 
\begin{equation}
D=\frac \kappa {2r}\left( \rho -<\rho >+2\Delta \right) .  \label{ten}
\end{equation}
A realistic requirement is that $\rho $ should be finite and positive. It
must decrease monotonically or stay constant for stability reasons, $\rho
^{\prime }\leq 0$. Then it is easily shown that $<\rho >^{\prime }\leq 0$
and $\rho \leq <\rho >$. Written in another way 
\begin{equation}
\frac{d\ln m}{d\ln r}\leq 3.  \label{eleven}
\end{equation}

Now, let us divide Eqs. (7) and (8): 
\begin{equation}
r\left( \ln \frac{z^{1/2}y^{\prime }}r\right) ^{\prime }=\frac{\rho -<\rho
>+2\Delta }{p+<\rho >/3}.  \label{twelve}
\end{equation}
There are two cases; $D\leq 0$ everywhere and $D>0$ somewhere. In the first
case the r.h.s. of Eq. (12) is bounded from above by zero. Perfect fluids ($%
\Delta =0$) form a subcase of this case. Anisotropic fluids with $\Delta
\leq 0$, which have radially dominated pressure, form another subcase. Even
when $\Delta >0$ in some regions there is a third subcase with $D\leq 0$. As
was mentioned in the introduction, a realistic $t$ satisfies the inequality $%
t\leq \varepsilon \rho $. Then $\Delta \leq t\leq \varepsilon \rho $ and a
sufficient condition for non-positive $D$ is 
\begin{equation}
\frac{d\ln m}{d\ln r}\leq \frac 3{1+2\varepsilon }.  \label{thirteen}
\end{equation}
When $\varepsilon =0$ we return to the isotropic case and Eq. (11) although $%
t=p$ instead of $t\leq 0$. Eq. (13) shows that when the slope of $m$ in a
logarithmic scale is not steep enough, no positive $\Delta $ is able to
ensure $D>0$. We shall show in the following section that models with $D\leq
0$ satisfy the Buchdahl bound on the redshift, while models with $D>0$ can
develop in principle bigger redshifts.

Anisotropic models are subjected to three field equations but possess five
characteristics $z,y,t,p,\rho $. Therefore, two of them or their
combinations must be given explicitly. Eq. (13) is important for models with
a given density profile. This profile should break at least once the
criterion in order to possibly achieve redshifts bigger than $2$. Another
possibility is to satisfy directly the condition $D>0$, which is more
general, but can be checked without solving the field equations only when $%
\Delta $ is the second given function.

When $D>0$ in some region, the following chain of inequalities holds 
\begin{equation}
\frac{\rho -<\rho >+2\Delta }{p+<\rho >/3}\leq \frac{2\left( t-p\right) }{%
p+<\rho >/3}\leq \frac{6t}{<\rho >}\leq 6\varepsilon .  \label{fourteen}
\end{equation}
The isotropic case is regained again when $\varepsilon =0$. Inserting the
bound from Eq. (14) into Eq. (12) and integrating from $r$ to the boundary
of the fluid sphere at $r=R$ yields 
\begin{equation}
y^{\prime }\geq A\left( R\right) \left( \frac rR\right) ^{6\varepsilon }%
\frac r{z^{1/2}},\quad A\left( r\right) =\frac{z^{1/2}y^{\prime }}r.
\label{fifteen}
\end{equation}
At $R$ the interior fluid solution should be matched to the exterior
Schwarzschild solution, which requires $z\left( R\right) =1-2M/R$. Here $%
M\equiv m\left( R\right) $ is the total mass of the fluid sphere. Eq. (9)
provides the inequality 
\begin{equation}
\frac{2m}r\geq \frac{2M}{R^3}r^2.  \label{sixteen}
\end{equation}

After these remarks, let us integrate Eq. (15) from the centre to the
boundary and take into account that $y\left( 0\right) \geq 0$. The result
reads 
\begin{equation}
\left( 1-\frac{2M}R\right) ^{1/2}\geq \frac M{2R^{3+6\varepsilon }}%
\int_0^{R^2}\frac{x^{3\varepsilon }dx}{\left( 1-\frac{2M}{R^3}x\right) ^{1/2}%
}.  \label{seventeen}
\end{equation}
Eq. (17) is a particular case of Eq. (66) from Ref. \cite{twentyone}. In
this paper polar Gaussian coordinates were used, the bound in Eq. (14) was
assumed to hold for some positive $\varepsilon $, not connected in general
with $t$, and the maximum of $2m/r$ was not obliged to be on the surface of
the configuration. Eq. (17) can be written as an inequality just for $\alpha
\equiv 2M/R$%
\begin{equation}
4\alpha ^{3\varepsilon }\left( 1-\alpha \right) ^{1/2}\geq \int_0^\alpha 
\frac{x^{3\varepsilon }dx}{\left( 1-x\right) ^{1/2}}.  \label{eighteen}
\end{equation}
This is the main inequality to be used for finding redshift bounds.

\section{Bounds and models}

Eq. (18) provides maximum values for the mass-radius ratio $\alpha $ and the
surface redshift 
\begin{equation}
s=\left( 1-\alpha \right) ^{-1/2}-1  \label{nineteen}
\end{equation}
both of which depend on $\varepsilon $. In the perfect fluid case formally $%
\varepsilon =0$ and Eq. (18) becomes 
\begin{equation}
3\left( 1-\alpha \right) ^{1/2}\geq 1.  \label{twenty}
\end{equation}
This gives the values found by Buchdahl, $\alpha _m=8/9$, $s_m=2$. When
anisotropy is present and $t$ is untied from $p$, $\varepsilon >0$ and there
is possibility for higher redshifts. In general, the integral on the r.h.s.
of Eq. (18) is expressed through the hypergeometric function. When $%
3\varepsilon $ is an integer it becomes a sum of different powers of $%
1-\alpha $. Thus, when DEC holds for $t$ ($\varepsilon =1$) we have 
\begin{equation}
\left( 1-\alpha \right) ^{1/2}\left[ \alpha +2\alpha ^3-\frac 17\left(
1-\alpha \right) ^3+\frac 35\left( 1-\alpha \right) ^2\right] \geq \frac{16}{%
35}  \label{twentyone}
\end{equation}
and the maximum bounds are $\alpha _m=0.974$ and $s_m=5.211$. These values
were found in Ref. \cite{twentyone}.

An useful relation which follows from the field equations is the
Tolman-Oppenheimer-Volkoff (TOV) equation \cite{six} 
\begin{equation}
p^{\prime }=-\left( \rho +p\right) \frac{1-z+\kappa pr^2}{2rz}+\frac{2\Delta 
}r.  \label{twentytwo}
\end{equation}
It shows that for realistic and, hence, finite at the centre $p,p^{\prime }$
we have $\Delta \left( 0\right) =0$. Furthermore, the radial pressure is
obliged to vanish at the boundary. If the more imperative SEC holds for $t$
, $\rho \geq p+2t$ gives $t\leq \frac 12\rho $ and even $t\left( 0\right)
\leq \frac 13\rho \left( 0\right) $. Let us accept the stronger bound, which
holds in the interior of perfect fluids, and call this a pseudoisotropic
case, with $t,p$ being of the same magnitude, although not equal. Then $%
\varepsilon =1/3$ and Eq. (18) becomes 
\begin{equation}
\left( 1-\alpha \right) ^{1/2}\left( 1+\frac 72\alpha \right) \geq 1,
\label{twentythree}
\end{equation}
which yields $\alpha _m=0.946$ and $s_m=3.310$. These values are still above
the Buchdahl ones.

The general value of $\varepsilon $ for SEC, however, is $1/2$ and this case
is the most realistic one. Now $3\varepsilon $ is not an integer, but the
integral in Eq. (18) is once more expressed in terms of elementary functions
and gives 
\begin{equation}
4\left( 1-\alpha \right) ^{1/2}\alpha ^{3/2}\geq \frac{2\alpha ^3+\alpha
^2-3\alpha }{4\sqrt{\alpha -\alpha ^2}}+\frac 38\arcsin \left( 2\alpha
-1\right) +\frac{3\pi }{16}.  \label{twentyfour}
\end{equation}
Computer calculations show that $\alpha _m=0.957$ and $s_m=3.842$. The last
number is the main result of this paper. It shows an almost double increase
in the Buchdahl bound when anisotropy is allowed and $t$ satisfies SEC. The
other assumptions made were $\rho >0,p\geq 0,\rho ^{\prime }\leq 0$ and the
inevitable $t\geq p$. Of course, $D>0$ should also be true, otherwise the
bound collapses to the Buchdahl's one. Thus, realistic anisotropic star
models can possess higher redshifts than the isotropic ones but it is
limited and never reaches infinity.

It is said sometimes for perfect fluids that the Buchdahl bound is optimal
because there is a model which saturates it. Eq. (12) shows that for such a
model $\rho =<\rho >$ which is possible only for constant $\rho $. Thus we
arrive at the Schwarzschild incompressible sphere with the central pressure
as a free parameter. Only the model with $p\left( 0\right) =\infty $
saturates the bound. A possible explanation is that $y\left( 0\right) \geq 0$
and the equality is required for saturation. This leads to singular metric
and it is probably induced only by singular central pressure. When the
pressure is required to satisfy SEC one has only $s_m=1/2$ \cite{thirteen}.
We leave aside the problem that constant density leads to infinite speed of
sound $v=\left( dp/d\rho \right) ^{1/2}$, while a realistic speed of sound
is bound by $1$, the speed of light in our units. In conclusion, there is no
realistic model, saturating the Buchdahl bound.

What is the situation for anisotropic models? According to Eq. (12) a
saturating model must satisfy 
\begin{equation}
\rho -<\rho >+2\Delta =6\varepsilon \left( p+\frac 13<\rho >\right) .
\label{twentyfive}
\end{equation}
However, this is not possible at the center because there the l.h.s.
vanishes, while the r.h.s. is positive. Again the explanation may be
connected with the requirement $p\left( 0\right) =\infty $ because of $%
y\left( 0\right) =0$. Then $\Delta \left( 0\right) <0$ and $D\left( 0\right)
<0$ since $t\left( 0\right) $ is finite, but this contradicts the nature of
the saturation ($D>0$).

One may ask for models which satisfy the limiting assumptions made during
the derivation of Eq. (14). Namely, $\rho =const=\rho _0$, $p=0$ and $%
t=\varepsilon \rho _0$. These are three conditions while only two of the
fluid's characteristics can be fixed beforehand. If we take the first two
conditions, we arrive at the Lemaitre model \cite{three,four,five}. Eq. (22)
then gives 
\begin{equation}
t=\frac{1-z}{4z}\rho _0.  \label{twentysix}
\end{equation}
Obviously $t$ is not as required. Using the bound on $t$ gives 
\begin{equation}
\alpha \leq \frac{4\varepsilon }{1+4\varepsilon }.  \label{twentyseven}
\end{equation}
This inequality yields $\alpha _m=0.8$, $s_m=1.236$ for DEC and $\alpha
_m=0.667$, $s_m=0.732$ for SEC. These values are higher than the ones of the
realistic Schwarzschild interior solution, but faraway from the absolute
bounds derived above. Two models possessing the values when DEC holds were
given recently \cite{seventeen} (models I and IV).

Most successful is the Bondi model \cite{thirteen} which also has a constant
density, but the second given function is the constant ratio $Q=\left(
p+2t\right) /\rho _0$. The region $Q\leq 1$ is studied, which is equivalent
to SEC. The redshift increases with $Q$ and numerical simulation gives $%
s_m=1.352$ for $Q=1$. Close to this result stands Example 2 from Ref. \cite
{eighteen} with $s=1.2$. It utilizes a nonlocal equation of state and a
density profile used in Refs. \cite{four,nine,fourteen}.

Finally, the conformally flat anisotropic models are worth being mentioned 
\cite{nine}. The vanishing of the Weyl tensor implies 
\begin{equation}
\frac{\kappa \Delta }r=-2\left( \frac m{r^3}\right) ^{\prime }.
\label{twentyeight}
\end{equation}
The two terms in $D$ are of the same nature, but do not compensate each
other. All characteristics of the solution depend only on the density
profile. $D$ is definitely positive, however, already the Buchdahl bound is
reached with infinite central radial pressure. This is analogous to the
behavior of the interior Schwarzschild solution which is the only
conformally flat solution in the perfect fluid case \cite{twentyfour}.

\section{Discussion}

The bounds found in this paper generalize the Buchdahl bound to anisotropic
star models and show that claims for arbitrarily large redshifts are not
realistic. The bounds are absolute, i.e. numbers depending on few simple
realistic requirements. They do not depend on the details of the mechanism
generating anisotropy, equations of state, central density and so on. We
have not explored also the consequences of the condition $0\leq v\leq 1$,
necessary for the causal behavior of the fluid. In the perfect fluid case a
lower bound $s_m=0.854$ was established heuristically when some of the above
mentioned features are taken into account \cite{twentyfive}. Probably, such
a study can be performed in the anisotropic case too. It may concern not
only the surface redshift but also the maximum masses and moments of inertia
of neutron stars.

\end{document}